\documentclass[preprint,showpacs,amsfonts,amsmath,amssymb]{revtex4}
\usepackage{graphicx}
\usepackage{dcolumn}
\usepackage{bm}
\usepackage{showlabels}

\begin{document}
\title{Quasinormal Modes of Scalar Field in Five-dimensional Lovelock Black Hole Spacetime}
\author{Juhua Chen$^{1,2}$} \email{jhchen@hunnu.edu.cn}
\author{Yongjiu Wang$^{1}$}
\affiliation{College of Physics and Information Science, Hunan
Normal University, Changsha, Hunan 410081, P. R. China.
\\Department of Physics \& Astronomy,
      University of Missouri, Columbia, MO 65211, USA.}
\begin{abstract}
In this paper using the third-order WKB approximation, a numerical
method devised by Schutz, Will and Iyer, we investigate the
quasinormal frequencies of the scalar field in the background of
five-dimensional Lovelock black hole. We find that the ultraviolet
correction to Einstein theory in the Lovelock theory makes the
scalar field decay more slowly and makes the scalar field oscillate
more quickly, and  the cosmological constant makes the scalar field
decay more slowly and makes the scalar field oscillate more slowly
in Lovelock black hole backgroud. On the other hand we also find
that quasinormal frequencies depend very weakly on the angular
quantum number $l$.

\end{abstract}
\maketitle
\section{Introduction}
The quasinormal modes\cite{Konoplya}, depending only on a black hole
parameters, are of great importance in gravitational-wave
astrophysics, and might be observed in existing or advanced
gravitational-wave detectors. Furthermore, black holes are often
used as a testing ground for ideas in quantum gravity, and their
quasinormal modes are obvious candidates for an interpretation in
terms of quantum levels\cite{Maggiore}. Because it is so important
for black hole physics and gravitational-wave astrophysics, there
are a lot of authors who are focus on the quasinormal modes of
matter fields in different black hole background in the past decade.
Such as: Quasinormal modes of black holes in anti-de Sitter
space\cite{Morgan}; the Dirac field quasinormal modes\cite{Sayan}
and the scalar field quasinormal modes\cite{Wang,Chakrabarti} in
different backgrouds. In recently some scholars investigated effects
of dark energy and dark matter on quasinormal modes\cite{He} and
some extended the investigation of the.quasinormal modes to higher
dimensional spacetimes\cite{Ortega}.

Lovelock\cite{Lovelock} extended the Einstein tensor, which is the
only symmetric and conserved tensor depending on the metric and its
derivatives up to the second order, to the most general tensor. They
obtained tensor is non linear in the Riemann tensor and differs from
the Einstein tensor only if the space-time has more than 4
dimensions. Therefore, the Lovelock theory is the most natural
extension of general relativity in higher dimensional space-times.
On the other hand, Lovelock theory resembles also string inspired
models of gravity as its action contains, among others, the
quadratic Gauss-Bonnet term, which is the dimensionally extended
version of the four-dimensional Euler density. This quadratic term
is present in the low energy effective action of heterotic string
theory\cite{Callan}. Since the Lovelock theory represents a very
interesting scenario to study how the physics of gravity results
corrected at short distance due to the presence of higher order
curvature terms in the action. C. Garraffo et al \cite{Garraffo}
gave a black hole solutions of this theory, and discussed how short
distance corrections to black hole physics substantially change the
qualitative features. And M. Aiello et al \cite{Aiello} presented
the exact five-dimensional charged black hole solution in Lovelock
gravity coupled to Born- Infeld electrodynamics. In their paper they
also investigated thermodynamical properties of lovelock black hole
spacetime. Further-more, M. H. Dehghani and R. Pourhasan
\cite{Dehghani} focused on the temperature of the uncharged black
holes of third order lovelock gravity and the entropy through the
use of first law of thermodynamics. They analyzed thermodynamical
stability  and found that there exists an intermediate
thermodynamically unstable phase for black holes with hyperbolic
horizon. R. A. Konoplya et al\cite{Abdalla} presented analysis of
the scalar perturbations in the background of Bauss-Bonnet black
hole spacetimes and its (in)stability in high
dimensions\cite{Roman}.

The aim of this paper is to study the quasinormal mode of a scalar
field in the Lovelock black hole spacetime in five-dimensional for
different angular quantum number $l$ by using the third-order WKB
approximation, a numerical method devised by Schutz, Will and Iyer
\cite{Schutz}. The paper is organized as follows: In section II we
will give a brief review on the Lovelock black hole spacetime in
five dimensions. In Section III a detail analysis on the quasinormal
mode of a scalar field in the Lovelock black hole spacetime in
five-dimensional is performed. In the last section a brief
conclusion is given.

\section{Lovelock Black hole spacetime in five dimensions}
The Lovelock Lagrangian density in $D$ dimensions is \cite{Lovelock}
\begin{eqnarray} \label{lov1}
L=\sum_{k=0}^{N} \alpha_{k} \lambda^{2(k-1) } L_{k},
\end{eqnarray}
where $N= \frac{D}{2}-1$ (for even $D$) and $N=\frac{D-1}{2}$ (for
odd $D$). In (\ref{lov1}), $\alpha _k$ and $\lambda $ are constants
which represent the coupling of the terms in the whole Lagrangian
and give the proper dimensions.

In  Eq. (\ref{lov1}) $L_{k}$ is
\begin{eqnarray} \label{lov2}
L_{k} = \frac {1}{2^{k}}
\sqrt{-g}\delta^{i_1...i_{2k}}_{j_1...j_{2k}} R^{j_{1} j_{2}}_{i_{1}
i_{2}}...\:R^{j_{2k-1}j_{2k}}_{i_{2k-1}i_{2k}},
\end{eqnarray}
where ${R^{\mu}\:_{\nu\rho\gamma}}$ is the Riemann tensor in $D$
dimensions,
$R^{\mu\nu}\:_{\rho\sigma}=g^{\nu\delta}\:R^{\mu}\:_{\delta\rho\sigma}$,
$g$ is the determinant of the metric $g_{\mu\nu}$ and
$\delta^{i_1...i_{2k}}_{j_1...j_{2k}}$ is the generalized Kronecker
delta of order $2k$ \cite{Misner}.
\bigskip

The Lagrangian up to order 2 are given by \cite{Lanczos}
\begin{eqnarray}
L_{0} &=& \sqrt{-g},\\
L_{1} &=& \frac{1}{2} \sqrt{-g} \delta^{i_{1} i_{2}}_{j_{1}
j_{2}}R^{j_{1} j_{2}}_{i_{1} i_{2}}= \sqrt{-g} R,\\
L_{2} &=& \frac{1}{4}\sqrt{-g} \delta^{i_{1} i_{2} i_{3}
i_{4}}_{j_{1} j_{2} j_{3} j_{4}} R^{j_{1} j_{2}}_{i_{1}
i_{2}}R^{j_{3} j_{4}}_{i_{3} i_{4}}
=\sqrt{-g}(R_{\mu\nu\rho\sigma}R^{\mu\nu\rho\sigma}-4
R_{\mu\nu}R^{\mu\nu}+R^2),
\end{eqnarray} where we recognize the
usual Lagrangian for the cosmological term, the Einstein-Hilbert
Lagrangian and the Lanczos Lagrangian \cite{Lanczos}, respectively.

For dimensions $D=5$ and $D=6$ the Lovelock Lagrangian is a linear
combination of the Einstein-Hilbert and Lanczos Lagrangian.

Hence, the geometric action is written as
\begin{eqnarray} \label{accion}
S=\int L d^Dx.
\end{eqnarray}

In this paper we only consider the spacetime in five dimensions, so
the Lagrangian is a linear combination of the Einstein-Hilbert and
the Lanczos ones, and the Lovelock tensor results
\begin{eqnarray}
{\cal
G}_{\mu\nu}=R_{\mu\nu}-\frac{1}{2}\:R\:g_{\mu\nu}+\Lambda\:g_{\mu\nu}
\end{eqnarray}
\begin{eqnarray}
-\alpha\:\{\frac{1}{2}\:g_{\mu\nu}\:(R_{\rho\delta\gamma\lambda}\:R^{\rho
\delta\gamma\lambda}-4\:R_{\rho\delta}\:R^{\rho\delta}+R^2)-
\end{eqnarray}
\begin{eqnarray} \label{love}
2\:R\:R_{\mu\nu}+4\:R_{\mu\rho}\:R^{\rho}_{\nu}+4\:R_{\rho\delta}
\:R^{\rho \delta}_{\mu
\nu}-2\:R_{\mu\rho\delta\gamma}\:R_{\nu}^{\rho\delta\gamma}\}.
\end{eqnarray}

The five-dimension Lovelock theory mainly corresponds to Einstein
gravity coupled to the dimensional extension of four dimensional
Euler density, that's to say, the theory referred as
Einstein-Gauss-Bonnet theory. The  spherically symmetric solution in
five dimensions take  as the follow form:
\begin{eqnarray} \label{metric}
ds^2=-N(r)dt^2+N^{-1}(r)dr^2+r^2d\Omega^{2}_{3},
\end{eqnarray}

where $d\Omega^{2}_{3}$ is the metric of a unitary 3-sphere, and
\begin{eqnarray} \label{solnueva2}
N(r)=\frac{4\alpha-4M+2r^2-\Lambda
r^4/3}{4\alpha+r^2+\sqrt{r^4+\frac{4}{3}\alpha\Lambda
r^4+16M\alpha}},
\end{eqnarray}
where  $M, \Lambda$  are  ADM mass, cosmological constant,
respectively, and $\alpha$ is the coupling constant of additional
term that presents the ultraviolet correction to Einstein theory.

\section{quasinormal mode of a scalar field in the Lovelock Black hole spacetime}
\begin{figure}[htbp]
\begin{center}
\includegraphics{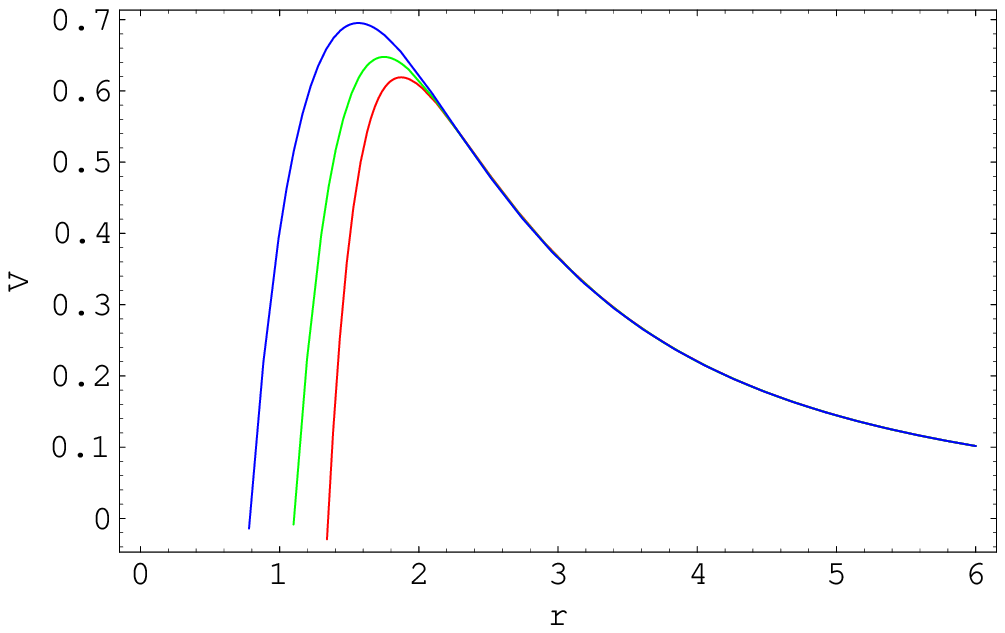}
\end{center}
\caption{The behavior of the effective potential $V(r)$ vs $r$ for
the Lovelock Black hole by fixed parameters $l=1,M=1,\Lambda=0.1$
and coupling constants $\alpha=0.1(red),0.4(yellow),0.7(blue)$.}

\begin{center}
\includegraphics{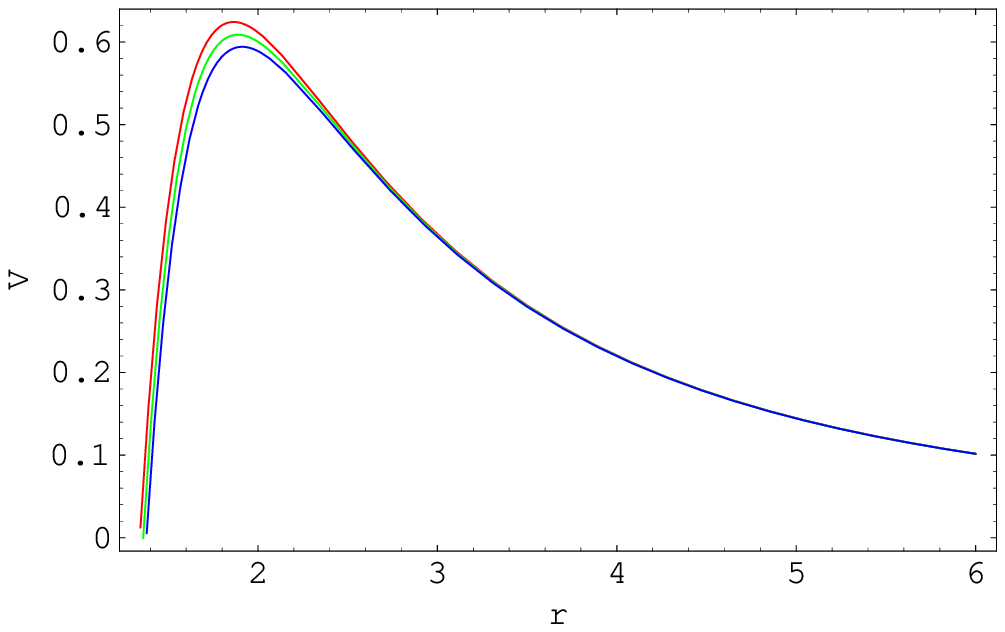}
\end{center}
\caption{The behavior of the effective potential $V(r)$ vs $r$ for
the Lovelock black hole by fixed parameters $l=1,M=1,\alpha=0.1$ and
cosmological constants $\Lambda=0(red),0.3(yellow),0.6(blue)$.}

\begin{center}
\includegraphics{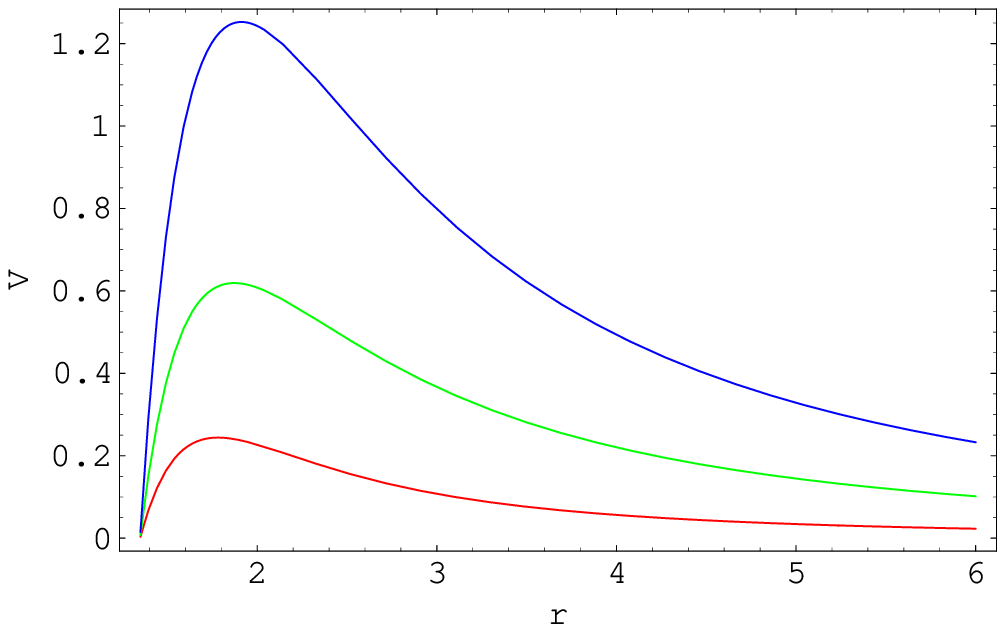}
\end{center}
\caption{The behavior of the effective potential $V(r)$ vs $r$ for
the Lovelock black hole  by fixed parameters
$M=1,\Lambda=\alpha=0.1$ and angular quantum numbers
$l=1(red),2(yellow),3(blue)$.}
\end{figure}

\begin{figure}[htbp]
\begin{center}
\includegraphics{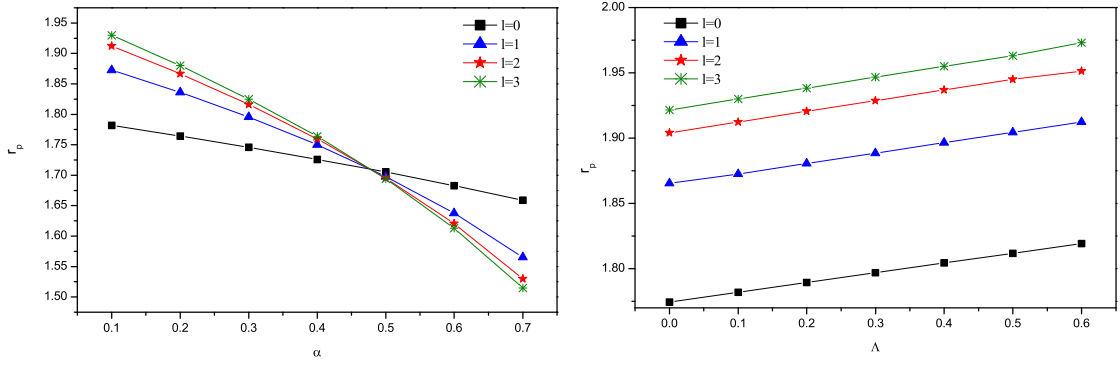}
\end{center}
\caption{The peak point ($r=r_p$)of the effective potential vs the
parameters of the Lovelock black hole for different angular quantum
numbers. The left corresponds to Fig.1 and the right corresponds to
Fig.2.}
\end{figure}

\begin{figure}[htbp]
\begin{center}
\includegraphics{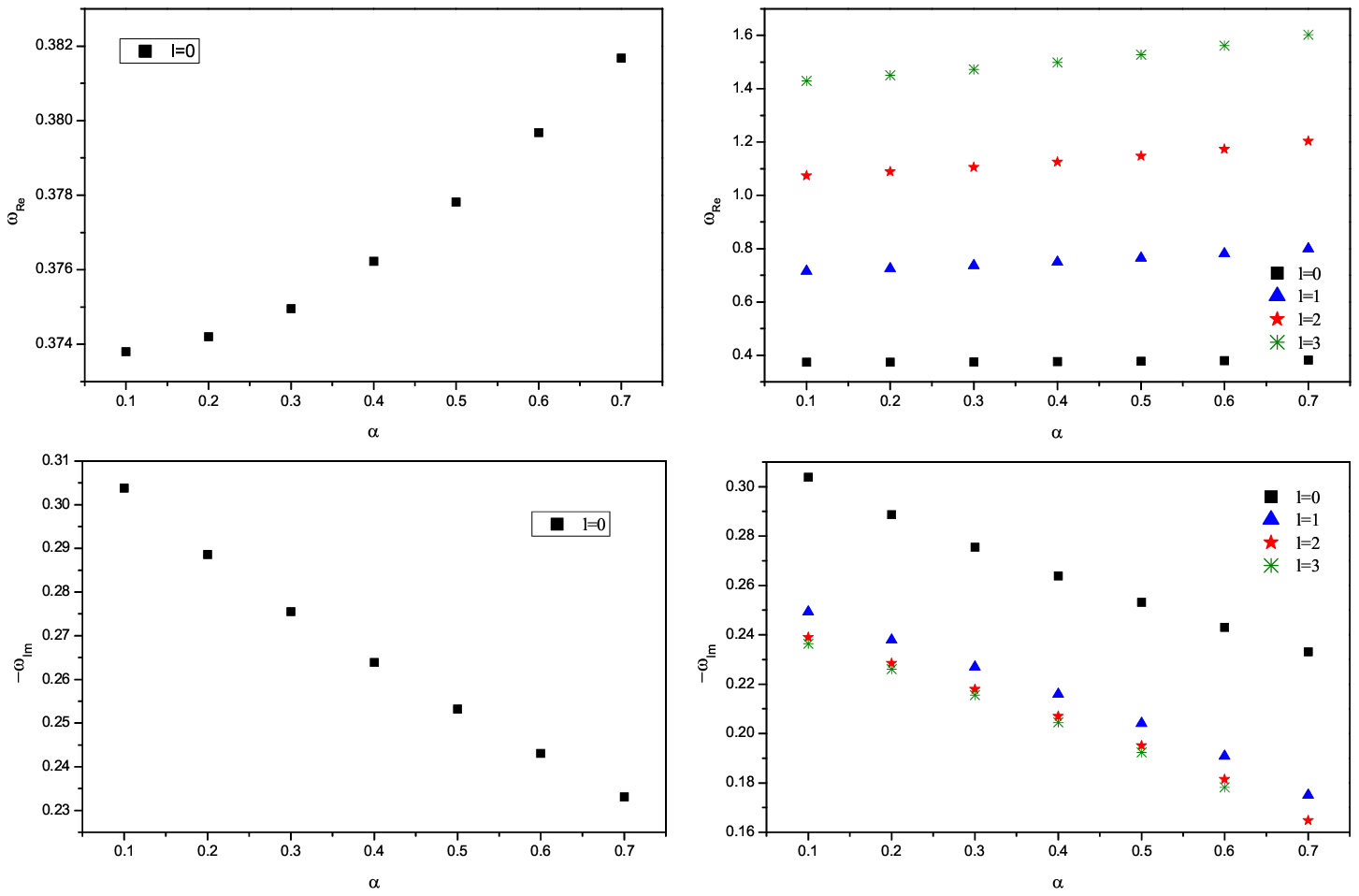}
\end{center}
\caption{Variation of the real parts (the above row) and imaginary
parts (the bottom row) of quasinormal frequencies of the scalar
field
 in  the Lovelock black hole spacetime with parameters $M=1,\Lambda=0.1$.}
\end{figure}

\begin{figure}[htbp]
\begin{center}
\includegraphics{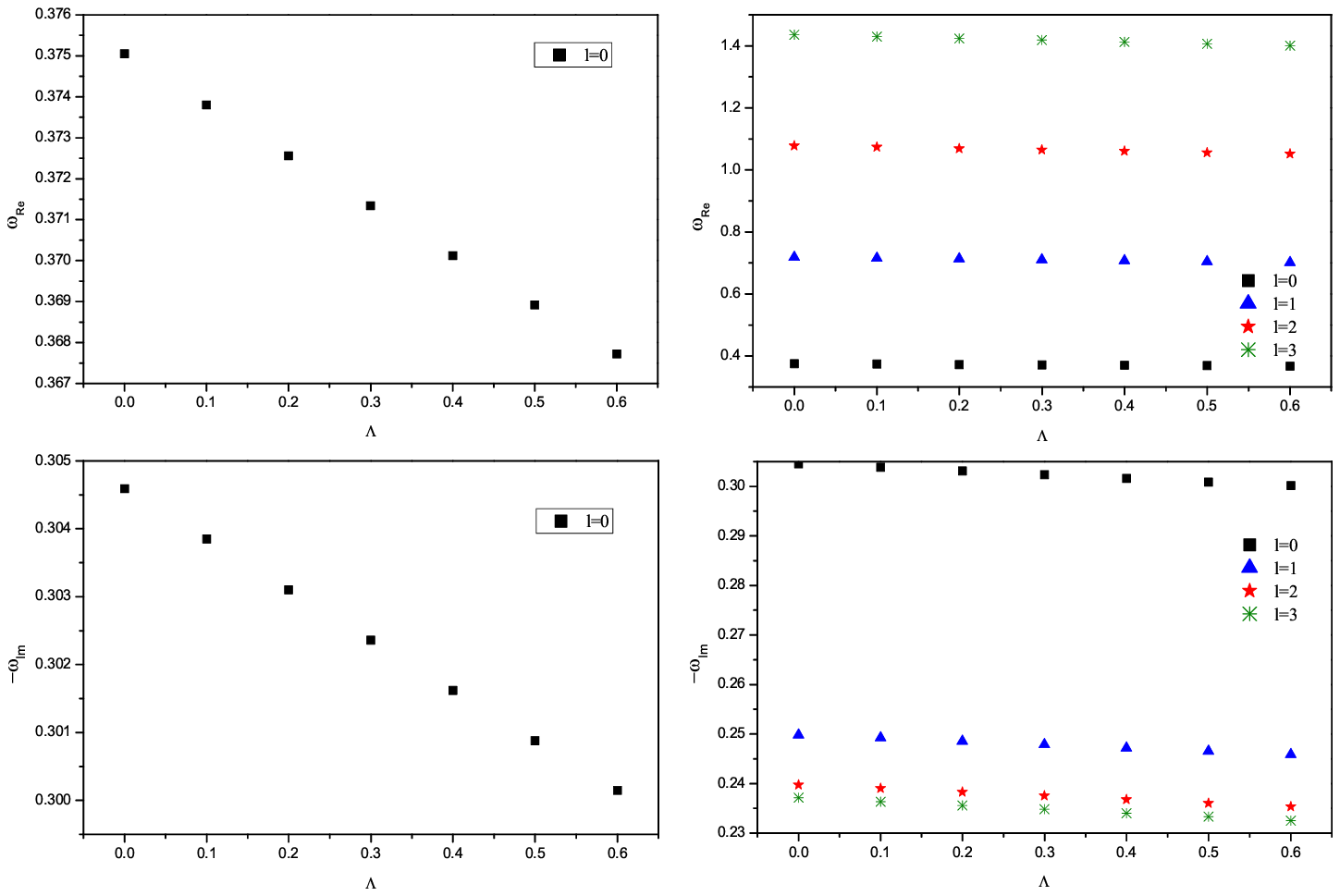}
\end{center}
\caption{Variation of the real parts (the above row) and imaginary
parts (the bottom row) of quasinormal frequencies of the scalar
field
 in  the Lovelock black hole spacetime with parameters $M=1,\alpha=0.1$.}
\end{figure}

The general perturbation equation for the massless scalar field in
the curve spacetime is given by
 \begin{eqnarray}
 \frac{1}{\sqrt{-g}}\partial_\mu(\sqrt{-g}g^{\mu\nu}
 \partial_\nu)\psi=0,\label{eq1}
 \end{eqnarray}
where $\psi$ is the scalar field.

Introducing the variables $\psi=\frac{e^{-i\omega
t}\Phi(r)}{r}Y(\theta,\varphi)$ and  $r_*=
\int{\frac{4\alpha-4M+2r^2-\Lambda
r^4/3}{4\alpha+r^2+\sqrt{r^4+\frac{4}{3}\alpha\Lambda
r^4+16M\alpha}}dr}$, and substituting Eq.(\ref{solnueva2}) into
Eq.(\ref{eq1}), we obtain a radial perturbation equation
\begin{eqnarray}
\frac{d^2\Phi(r)}{dr_*^2}+(\omega^2-V(r))\Phi(r)=0,\label{e3}
\end{eqnarray}
where
\begin{eqnarray}
V(r)&=& \frac{4\alpha-4M+2r^2-\Lambda
r^4/3}{4\alpha+r^2+\sqrt{r^4+\frac{4}{3}\alpha\Lambda
r^4+16M\alpha}}[\frac{l(l+2)}{r^2}  \nonumber \\
&+&\frac{3}{4r^2}\frac{4\alpha-4M+2r^2-\Lambda
r^4/3}{4\alpha+r^2+\sqrt{r^4+\frac{4}{3}\alpha\Lambda r^4+16M\alpha}}  \nonumber \\
&+&\frac{1}{4\alpha}(3-\frac{3+4\alpha M}{\sqrt{9r^4+12\alpha\Lambda
r^4+144M\alpha}}r^2)] \label{v}.
 \end{eqnarray}

It is obvious that the effective potential $V$ depends only on the
value of $r$, angular quantum number $l$, ADM mass $M$, cosmological
constant $\Lambda$ and coupling constant $\alpha$, respectively.
Fig.1 and the left one of Fig.4 show the variation of the effective
potential and its peak point $r_{p}$ with respect to the coupling
constant $\alpha$. From these two figures we can find that the peak
value of potential barrier gets lower and the location of the peak
($r=r_p$) moves along the right when the coupling constant $\alpha$
decreases. In Fig.2 and the right one of Fig.4 we give the variation
of the effective potential and the its peak point $r_{p}$ with
respect to the cosmological constant $\Lambda$. On the other side,
from these two figures we can find that the peak value of potential
barrier gets lower and the location of the peak ($r=r_p$) moves
along the right when the coupling constant $\Lambda$ increases,
which is different from the coupling constant $\alpha$. But from
Fig.3 we can see that the peak value of potential barrier gets upper
and the location of the peak point ($r=r_p$) moves along the right
when the angular quantum number $l$ increases.

From effective potential $V(r)$, i.e., Eq.(\ref{v}) and Fig.1,2, we
find that the quasinormal frequencies depend on the coupling
constant $\alpha$ and the cosmological constant $\Lambda$. In this
paper, we plan to investigate the relationship between the
quasinormal mode and the coupling constant $\alpha$ and the
cosmological constant $\Lambda$, respectively. For convenience we
take $M=1$ in our calculation. In order to evaluate the quasinormal
frequencies for the massless scalar field in the Lovelock black hole
spacetime (\ref{metric}), we use the third-order WKB approximation,
a numerical method devised by Schutz, Will and Iyer \cite{Schutz}.
This method has been used extensively in evaluating quasinormal
frequencies of various black holes because of its considerable
accuracy for lower-lying modes. In this approximate method, the
formula for the complex quasinormal frequencies $\omega$ is
\begin{eqnarray}
\omega^2=[V_0+(-2V^{''}_0)^{1/2}\Lambda]-i(n+\frac{1}{2})(-2V^{''}_0)^{1/2}(1+\Omega),
\end{eqnarray}
where
\begin{eqnarray}
\Lambda&=&\frac{1}{(-2V^{''}_0)^{1/2}}\left\{\frac{1}{8}\left(\frac{V^{(4)}_0}{V^{''}_0}\right)
\left(\frac{1}{4}+N^2\right)-\frac{1}{288}\left(\frac{V^{'''}_0}{V^{''}_0}\right)^2
(7+60N^2)\right\},\\
\Omega&=&\frac{1}{(-2V^{''}_0)^{1/2}}\bigg\{\frac{5}{6912}
\left(\frac{V^{'''}_0}{V^{''}_0}\right)^4 (77+188N^2)\nonumber\\&-&
\frac{1}{384}\left(\frac{V^{'''^2}_0V^{(4)}_0}{V^{''^3}_0}\right)
(51+100N^2)
+\frac{1}{2304}\left(\frac{V^{(4)}_0}{V^{''}_0}\right)^2(67+68N^2)
\nonumber\\&+&\frac{1}{288}
\left(\frac{V^{'''}_0V^{(5)}_0}{V^{''^2}_0}\right)(19+28N^2)-\frac{1}{288}
\left(\frac{V^{(6)}_0}{V^{''}_0}\right)(5+4N^2)\bigg\},
\end{eqnarray}
and
\begin{eqnarray}
N=n+\frac{1}{2},\;\;\;\;\;
V^{(n)}_0=\frac{d^nV}{dr^n_*}\bigg|_{\;r_*=r_*(r_{p})}.
\end{eqnarray}

Substituting the effective potential (\ref{v}) into the formula
above, we can obtain the quasinormal frequencies of the scalar field
in the background of five-dimensional Lovelock black hole. Fig.5 and
Table.I show the real and imagine parts of quasinormal frequencies
for the scalar field with the variation of coupling constant
$\alpha$ and angle quantum number $l$. By analyzing these data and
curves, we can find that, when the coupling constant $\alpha$ (i.e.
the additional term presents the ultraviolet correction to Einstein
theory) increases,  the real part quasinormal frequencies of the
scalar field increases, while the imaginary part decreases, which
means that the ultraviolet correction makes the scalar field decay
more slowly and makes the scalar oscillate more quickly. Fig.6 and
Table.II show the real and imagine parts of quasinormal frequencies
for the scalar field with the variation of the cosmological constant
$\Lambda$ and angle quantum number $l$. Base on the data, we can
make a conclusion that, when the cosmological constant $\Lambda$
increases, the real part and the imaginary part of quasinormal
frequencies of the scalar field decreases, that's to say which means
that the cosmological constant makes the scalar field decay more
slowly and makes the scalar oscillate more slowly.

Moreover, The Re($\omega$) increases (decreases the oscillatory time
scale) and the Im($\omega$) decreases (increases the damping time
scale) as the angular quantum number $l$ increases for fixed n,
quasinormal frequencies depend very weakly on the angular quantum
number $l$, which is the same as Jing's\cite{Jing}.

\begin{table}[h1]
\caption{Quasinormal frequencies of the scalar field in  the
Lovelock black hole spacetime with parameters $M=1,\Lambda=0.1$ and
$n=0$.}
\begin{tabular}[b]{c|c|c|c|c}
 \hline \hline
 $\alpha$ &$\omega\;(l=0)$& $\omega \;(l=1)$ &$\omega \;(l=2)$ &$\omega \;(l=3)$  \\ \hline
0.1&0.373800-0.303847i&0.715847-0.249250i&1.07354-0.238997i&1.43010-0.236352i
 \\
0.2&0.374204-0.288632i&0.725245-0.237874i&1.08845-0.228468i&1.45024-0.226095i
 \\
0.3&0.374958-0.275545i&0.736564-0.226962i&1.10557-0.217960i&1.47295-0.215606i
\\
0.4&0.376224-0.263901i&0.749615-0.215928i&1.12509-0.207052i&1.49867-0.204530i
\\
0.5&0.377815-0.253200i&0.764359-0.204153i&1.14737-0.195167i&1.52804-0.192338i
\\
0.6&0.379676-0.243030i&0.780916-0.190888i&1.17301-0.181496i&1.56212-0.178243i
\\
0.7&0.381678-0.233081i&0.799620-0.175064i&1.20317-0.164778i&1.60272-0.161014i
\\
\hline \hline
\end{tabular}
 \end{table}

\begin{table}[h2]
\caption{Quasinormal frequencies of the scalar field  in  the
Lovelock black hole spacetime with parameters $M=1,\alpha=0.1$ and
$n=0$.}
\begin{tabular}[b]{c|c|c|c|c}
 \hline \hline
 $\Lambda$ &$\omega\;(l=0)$& $\omega \;(l=1)$ &$\omega \;(l=2)$ &$\omega \;(l=3)$  \\ \hline
  0&0.375051-0.304596i&0.718738-0.249821i&1.07808-0.239750i&1.43618-0.237140i
\\
0.1&0.373800-0.303847i&0.715847-0.249250i&1.07354-0.238997i&1.43010-0.236352i
 \\
0.2&0.372561-0.303102i&0.712923-0.248564i&1.06905-0.238251i&1.42409-0.235571i
 \\
0.3&0.371335-0.302359i&0.710035-0.247882i&1.06462-0.237511i&1.41815-0.234797i
\\
0.4&0.370118-0.301619i&0.707182-0.247203i&1.06024-0.236776i&1.41229-0.234030i
\\
0.5&0.368915-0.300883i&0.704363-0.246530i&1.05591-0.236049i&1.40650-0.233271i
\\
0.6&0.367723-0.300150i&0.701578-0.245860i&1.05164-0.235327i&1.40078-0.232518i
\\

\hline \hline
\end{tabular}
 \end{table}
\section{conclusions}
 Using the third-order WKB approximation, a numerical method devised
by Schutz, Will and Iyer, we obtained the quasinormal frequencies of
the scalar field in the background of five-dimensional Lovelock
black hole in further detail. we can find that the ultraviolet
correction to Einstein theory in the Lovelock theory makes the
scalar field decay more slowly and makes the scalar field oscillate
more quickly, and the cosmological constant makes the scalar field
decay more slowly and makes the scalar field oscillate more slowly
in Lovelock black hole backgroud. At the same time we also find that
quasinormal frequencies depend very weakly on the angular quantum
number $l$.
\section{Acknowledgments}
J.H. Chen is supported by National Natural Science Foundation of
China(Grant:10873004), Scientific Research Fund of Hunan Provincial
Education Department(Grant:08B051), program for excellent talents in
Hunan Normal University and State Key Development Program for Basic
Research Program of China (Grant: 2003CB716300).


\end{document}